\documentclass[preprint, showpacs,preprintnumbers,amsmath,amssymb,nofootinbib]{revtex4}
\usepackage{mathrsfs}
\usepackage{amsmath}
\usepackage{amsfonts}
\usepackage{latexsym}
\usepackage{amsfonts}
\usepackage{graphicx}
\usepackage{epsf}
\usepackage{dcolumn}
\usepackage{bm}
\usepackage{subfigure}

\newcommand{\bea}[1]{\begin{eqnarray}\label{#1}}
\newcommand{\eea}{\end{eqnarray}}

\makeatletter

\newcommand{\Rmnum}[1]{\expandafter\@slowromancap\romannumeral #1@}
\makeatother

\def\gsim{ \lower .75ex \hbox{$\sim$} \llap{\raise .27ex \hbox{$>$}} }
\def\lsim{ \lower .75ex \hbox{$\sim$} \llap{\raise .27ex \hbox{$<$}} }

\begin{document}
 \title{G\"{o}del-type universes in $f(T)$ gravity}

\author{ Di Liu, Puxun Wu 
and Hongwei Yu  }
\address
{Center of Nonlinear Science and Department of Physics, Ningbo
University,  Ningbo, Zhejiang, 315211 China }

\begin{abstract}
 The issue of causality in $f(T)$ gravity is investigated
 by examining the possibility of existence of the closed timelike
curves in the G\"{o}del-type metric.  By assuming a perfect fluid as
the matter source, we find that  the fluid must have an equation of
state parameter greater than minus one  in order to allow the
G\"{o}del solutions to exist,  and furthermore the critical radius
$r_c$, beyond which the causality is broken down, is finite and it
depends on both matter and gravity. Remarkably, for certain $f(T)$
models, the perfect fluid that allows the G\"{o}del-type solutions
can even be normal matter, such as pressureless matter or radiation.
However, if the matter source is a special scalar field rather than
a perfect fluid, then $r_c\rightarrow\infty$ and  the causality
violation  is  thus avoided.

\end{abstract}
\maketitle

\section{Introduction}\label{sec1}
 General relativity (GR) is established in the framework of the Levi-Civita connection,
therefore there is only curvature rather than torsion in the spacetime. On the other hand,
one can also introduce other connections, such as the Weitzenb\"{o}ck connection, into the same
spacetime where only torsion is reserved.
Thus, there is no such a thing as curvature or torsion of
spacetime, but only curvature or torsion of connection. Basing on
the Weitzenb\"{o}ck connection,  Einstein~\cite{Einstein} introduced
firstly the Teleparallel  Gravity (TG) in his endeavor to unify
gravity and electromagnetism with the introduction of a tetrad
field.  TG can,  as is well known,  show up as a theory completely
equivalent to GR since the difference between their actions (the
actions of TG  and GR are the torsion scalar $T$ and Ricci scalar
$R$, respectively) is just a derivative term~\cite{FNGtnb, FNGtn1, FNGtn2, FNGtn3, FNGtn4, FNGtne}.

Recently, a modification of TG, called  $f(T)$
theory~\cite{Bengochea2009, Ferraro2007,  Linder, Zheng2011,
Ferraro2008, Ferraro2011, pwhy2011, Wu2011,Bamba2011,Wu2010a,Ben2011ab,Wu2010b, Zhang2011bb, FTbe, FT1,
FT2, FT3, FT4, FT5, FT6, FT7,FT8,FT9, FT10, FT11, FT12, FT13, FT14, FT15, FT16,FT17, FT18, FT19, FT20,
FT21, FT22, FT23,FT24,FT25, FT26, FT27, FT28, FT29, FT30, FT31, FT32, FT33, FT34, FT35, FT36, FT37,
FT38, FTed, Ferraro2011a, Li2011aa,Li2011bb,Li2011b, LiM2011, Miao2011}, has spurred an increasing deal of
attention, as it can explain the present accelerated cosmic
expansion discovered  from observations (the Type Ia
supernova~\cite{R98, P99}, the cosmic microwave background
radiation~\cite{Spa, Spb}, and the large scale structure ~\cite{T2004, E2005},
etc.) without the need of dark energy.  $f(T)$ theory is obtained by
generalizing the action $T$ of TG to an arbitrary function $f$ of
$T$, which is very analogous  to  $f(R)$ theory~(see
\cite{FeNoj08,FeSot10,Felice2010,FeNoj11,FebCli} for recent review) where the action $R$ of GR is
generalized to be $f(R)$. An advantage of $f(T)$ theory is that its
field equation is only second order, while in $f(R)$ gravity it is
forth order.

It has been found that $f(T)$ theory can give an inflation without
an inflaton~\cite{Ferraro2007, Ferraro2008},  avoid  the big bang
singularity problem in the standard cosmological
model~\cite{Ferraro2011}, realize the crossing of phantom divide
line for the effective equation of state~\cite{Wu2011, Bamba2011},
and yield an usual early cosmic evolution~\cite{Wu2010b, Zhang2011bb}.  But, at
the same, this theory lacks the local Lorentz
invariance~\cite{Li2011aa, Li2011bb}, and  this results in the appearance of
extra degrees of freedom~\cite{LiM2011}, the broken down of the
first law of black hole thermodynamic~\cite{Miao2011}, and  the
problem in cosmic large scale structure~\cite{Li2011b}.

In this paper, we plan to study the causality issue of $f(T)$ theory by examining the possibility of existence of the closed
timelike curves  in
the G\"{o}del spacetime~\cite{Godel}. The G\"{o}del metric  is the first cosmological solution with rotating matter to the
Einstein equation in GR.  Since the G\"{o}del solution is very convenient for studying whether the closed timelike curves exist, it has been
used widely to test the causality issue.
For example, G\"{o}del found that the closed timelike solution cannot be excluded in GR,  assuming a cosmological constant or a perfect fluid  with its pressure equal to the energy density. G\"{o}del's work has been generalized to include other matter sources, such as, the vector field~\cite{Sombe, SomRe79, SomRay80}, scalar field~\cite{Hiscockbe, HisCha, HisPan}, spinor field~\cite{Villalbabe, VilPim,VilKre,VilLea,VilHered, Reboucas1983} and tachyon field~\cite{Reboucas}. In addition, the G\"{o}del-type universes~\cite{mjrteibe,mjrtei1,mjrteied, Reboucas1983} have also been studied in the framework of other theories
of gravitation, such as TG~\cite{TGGod}, $f (R)$ gravity~\cite{RebClif05e, Reb09b1, Reb10ed} and string-inspired gravitational theory~\cite{stri,barrow1998}.

Here,  assuming that the matter source is the perfect fluid or a scalar field, we aim to find out the  condition for non-violation of causality in $f(T)$ gravity. The paper is organized as follows.  We give, in Sec. II, a brief review of  $f(T)$ theory and  the vierbein of a general cylindrical symmetry metric in Sec.III.   The  G\"{o}del-type universe in $f(T)$ theory is discussed in Sec. IV.  With an assumption of different matter sources,  we investigate the issue of causality in Sec. V.  Finally,  we present our conclusions in Sec. VI.

\section{ $f(T)$ gravity}\label{ftgravi}

In this section, we give a brief view of  $f(T)$ gravity. We use the Greek alphabet ($\mu $, $\nu $,
$\cdots$= 0, 1, 2, 3) to denote tensor indices, that is, indices related to
spacetime, and middle part of the Latin alphabet ($i$, $j$, $\cdots$= 0, 1, 2, 3) to
denote tangent space (local Lorentzian) indices. TG, instead of using the metric tensor,  uses tetrad, $e_{\mu }^{i}$ or $e_{i}^{\mu }$ (frame or coframe), as   the dynamical object.  The
relation between frame and coframe is
\begin{equation}\label{tetradralation}
e_{i}^{\mu }e_{\mu }^{j}=\delta _{i}^{j}\;,\qquad e_{i}^{\mu }e_{\nu
}^{i}=\delta _{\nu }^{\mu },\end{equation}
and the relation between  tetrad and metric tensor is
\begin{equation}\label{vierbien}
g_{\mu \nu }=e_{\mu }^{i}e_{\nu }^{j}\eta _{ij}\;,\qquad \eta _{ij}=e_{i}^{\mu }e_{j}^{\nu }g_{\mu
\nu }\;,
\end{equation}
where $\eta _{ij}=diag(1,-1,-1,-1)$ is the Minkowski metric.

Different from GR, the Weitzenb\"{o}ck connection is used in TG
\begin{equation}\label{conection}
{\Gamma }_{\mu \nu }^{\lambda }=e_{i}^{\lambda }\partial _{\nu }e_{\mu
}^{i}=-e_{\mu }^{i}\partial _{\nu }e_{i}^{\lambda }\;.
\end{equation}
As a result, the convariant derivative, denoted by $D_{\mu }$, satisfies:
\begin{equation}\label{drive}
D_{\mu }e_{\nu }^{i}=\partial _{\mu }e_{\nu }^{i}-\Gamma _{\nu \mu
}^{\lambda }e_{\lambda }^{i}=0\;.
\end{equation}
To describe the difference between Weitzenb\"{o}ck and Levi-Civita connections, a contorsion tensor
$K_{\;\;\mu \nu }^{\rho }$ needs to be introduced:
\begin{equation}\label{K}
K_{\;\;\mu \nu }^{\rho }\equiv \Gamma _{\mu \nu }^{\rho }-\overset{\circ }{\Gamma }{}_{\mu \nu }^{\rho
}=\frac{1}{2}(T_{\mu }{}^{\rho }{}_{\nu }+T_{\nu
}{}^{\rho }{}_{\mu }-T_{\;\;\mu \nu }^{\rho })\;.
\end{equation}
Here $T_{\;\;\mu \nu }^{\rho }$ is the torsion tensor
\begin{equation}\label{T}
T_{\;\;\mu \nu }^{\rho }={\Gamma }_{\nu \mu }^{\rho }-{\Gamma }_{\mu \nu
}^{\rho }=e_{i}^{\rho }(\partial _{\mu }e_{\nu }^{i}-\partial _{\nu }e_{\mu
}^{i})\;,
\end{equation}
and $\overset{\circ }{\Gamma }{}_{\mu \nu }^{\rho }$ denotes the
Levi-Civita connection
\begin{equation}
\overset{\circ }{\Gamma }{}_{\mu \nu }^{\rho }=\frac{1}{2}g^{\rho \sigma
}(\partial _{\mu }g_{\sigma \nu }+\partial _{\nu }g_{\sigma \mu }-\partial
_{\sigma }g_{\mu \nu }).
\end{equation}
By defining the super-potential $S_{\sigma }^{\;\;\mu \nu }$
\begin{equation}\label{S}
S_{\sigma }^{\;\;\mu \nu }\equiv K_{\;\;\;\;\sigma }^{\mu \nu
}+\delta _{\sigma }^{\mu }T_{\;\;\;\;\;\alpha }^{\alpha \nu }-\delta _{\sigma
}^{\nu }T_{\;\;\;\;\;\alpha }^{\alpha \mu }\;,
\end{equation}
we obtain the torsion scalar $T$
\begin{equation}\label{scalarT}
T\equiv \frac{1}{2}S_{\sigma }^{\;\;\mu \nu }T_{\;\;\mu \nu }^{\sigma }=
\frac{1}{4}T^{\alpha \mu \nu }T_{\alpha \mu \nu }+\frac{1}{2}T^{\alpha \mu
\nu }T_{\nu \mu \alpha }-T_{\alpha \mu }^{\;\;\;\;\alpha }T_{\;\;\;\;\;\nu }^{\nu \mu
}\;.
\end{equation}
In TG, the Lagrangian density is given by:
\begin{equation}
L_{T}=\frac{eT}{2\kappa ^{2}}\;,
\end{equation}
where, $e=\det (e_\mu^{i})=\sqrt{-g}\;, \kappa ^2{\equiv }8\pi G$.  Generalizing $T$ to be an arbitrary function $f$ of $T$ in the above expression, we obtain the Lagrangian
density of  $f(T)$ theory
\begin{equation}\label{f1}
L_{T}=\frac{ef(T)}{2\kappa ^2}\;.
\end{equation}
Adding a matter Lagrangian density $L_M$ to Eq.~(\ref{f1}), and
varying the action with respect to the vierbein, one finds the
following field equation of $f(T)$ theory:
\begin{eqnarray}\label{motion1}
&&[e^{-1}\partial_\mu (ee^\rho_i S^{\;\;\nu\mu}_\rho)-e^\lambda_i S^{\rho\mu\nu} T_{\rho\mu\lambda}]f_T(T)
+e^\rho_i S^{\;\;\nu\mu}_\rho \partial_\mu (T)f_{TT}(T)\\ \nonumber
&&+\frac{1}{2}e^\nu_i f(T)=\kappa^2 e^\rho_i \overset{em}{T}{}^{\;\;\nu}_{\rho}.
\end{eqnarray}
Here $f_T= df(T)/dT$, $f_{TT}= d^2f(T)/dT^2$,  and  $\overset{em}{T}{}^\nu_{\rho}$ is the matter energy-momentum tensor. In a coordinate system, this field equation can be rewritten as
\begin{eqnarray}\label{motion2}
A_{\mu\nu}f_T(T)+S^{\;\;\;\;\;\sigma}_{\nu\mu}(\nabla_\sigma T)f_{TT}(T)+\frac{1}{2}g_{\mu\nu} f(T)=\kappa^2\overset{em}{T}{}_{\mu\nu}\;,
\end{eqnarray}
where
\begin{eqnarray}
&&A_{\mu \nu }=g_{\sigma\mu}e^i_\nu[e^{-1}\partial_\xi(ee^\rho_iS^{\;\;\sigma\xi}_\rho)-e^\lambda_iS^{\rho\xi\sigma} T_{\rho\xi\lambda}]\\ \nonumber
&&\qquad=G_{\mu \nu }-\frac{1}{2}g_{\mu \nu }T=-\nabla^\sigma S_{\nu\sigma\mu }
-S_{\;\;\;\;\mu }^{\rho\lambda }K_{\lambda \rho \nu }\;,
\end{eqnarray}
$G_{\mu\nu}$ is the Einstein tensor, and $\nabla_{\sigma}$ is the covariant derivative associated with the  Levi-Civita connection.
The trace of Eq.~(\ref{motion1}) or (\ref{motion2}), which can be used to
simplify and constrain the field equation, can be expressed as
\begin{eqnarray}\label{trace}
-[2e^{-1}\partial_\sigma (eT^{\;\;\rho\sigma}_\rho)+T]f_T(T)+S^{\;\;\rho\sigma}_\rho (\partial_\sigma T)f_{TT}(T)+2f(T)=\kappa^2\overset{em}{T}\;,
\end{eqnarray}
where, $\overset{em}{T}=\overset{em}{T}{}^{\mu}_{\;\;\mu}=g^{\mu \nu }\overset{em}{T}{}_{\mu \nu }$ is the trace
of the energy-momentum tensor.
Clearly, in the case of TG, $f(T)=T$, and Eq.~(\ref{trace}) reduces to
\begin{equation}\label{trTEGR}
T-2e^{-1}\partial _{\sigma }(eT_{\rho }^{\;\;\rho \sigma })=\kappa^2\overset{em}{T}\;,
\end{equation}
which shows an equivalence between GR and TG since \begin{equation}\label{T+R}
-R=T-2e^{-1}\partial _{\sigma }(eT_{\rho }^{\;\;\rho \sigma })\;.
\end{equation}

\section{ vierbein for cylindrical symmetry metric }
Since the G\"{o}del-type metric  is usually expressed
in cylindrical coordinates $[(r,\phi,z)]$, we consider a general cylindrical symmetry metric
\begin{eqnarray}\label{ds21}
ds^2=dt^2+2H(r)dtd\phi-dr^2-G(r)d\phi^2-dz^2\;,
\end{eqnarray}
where $H$ and $G$ are  the arbitrary  functions of $r$. This metric
can be re-expressed in the following form
\begin{eqnarray}
ds^2=[dt+H(r)d\phi]^2-D^2(r)d\phi^2-dr^2-dz^2\;,
\end{eqnarray}
where
\begin{eqnarray}
D(r)=\sqrt{G(r)+H^2(r)}\;.
\end{eqnarray}
Since the local Lorentz invariance is violated in $f(T)$ theory and the
vierbein have  six  degrees of freedom more than the metric, one
should be careful in choosing a physically reasonable tetrad in
terms of Eq.(\ref{vierbien}). Here,  we choose the tetrad anstaz  of
the cylindrical symmetry metric to be:
\begin{eqnarray}\label{tet}
e^i_\mu\equiv \left(\begin{array}{cccc}
1&0&H&0\\0&1&0&0\\0&0&D&0\\0&0&0&1
\end{array}\right) \;,\;\;\;
e^\mu_i\equiv\left(\begin{array}{cccc}
1&0&-\frac{H}{D}&0\\0&1&0&0\\0&0&\frac{1}{D}&0\\0&0&0&1\\
\end{array}\right)\;.
\end{eqnarray}
Using Eqs.~(\ref{conection}--\ref{scalarT}), one can find that the
Weitzenb\"{o}ck invariant  $T$ is
\begin{eqnarray}\label{T}
T=\frac{1}{2}\left(\frac{H'}{D}\right)^2\;,
\end{eqnarray}
where a prime presents a derivative with respect to $r$.

Substituting the vierbein given in Eq.~(\ref{tet}) into  Eq.~(\ref{motion2}), we obtain the following non-zero components of the $f(T)$ field equation:
\\$\nu=0, i=0$
\begin{eqnarray}\label{ft00}
\bigg(T-\frac{D''}{D}+\frac{HT'}{2H'}\bigg)f_T(T)+\bigg(\frac{HT}{H'}-\frac{D'}{D}\bigg)T'f_{TT}(T) +\frac{1}{2}f(T)=\kappa^2\overset{em}{T}{}^0_{\;\;0}
\end{eqnarray}
$\nu=0, i=2$
\begin{eqnarray}\label{ft02}
\bigg(HT+\frac{T'D^2}{2H'}\bigg)f_T(T)+\frac{T'H'}{2}f_{TT}(T)-\frac{H}{2}f(T)=\kappa^2\bigg(\overset{em}{T}{}^0_{\;\;2}-
H\overset{em}{T}{}^0_{\;\;0}\bigg)\;,
\end{eqnarray}
$\nu=1, i=1$
\begin{eqnarray}\label{ft11}
-T f_T(T)+\frac{1}{2}f(T)=\kappa^2\overset{em}{T}{}^1_{\;\;1}\;,
\end{eqnarray}
$\nu=2, i=0$
\begin{eqnarray}\label{ft20}
T'\bigg[\frac{1}{2H'}f_T(T)+\sqrt{\frac{T}{2}}f_{TT}(T)\bigg]=\kappa^2\overset{em}{T}{}^2_{\;\;0}\;,
\end{eqnarray}
$\nu=2, i=2$
\begin{eqnarray}\label{ft22}
-Tf_T(T)+\frac{1}{2}f(T)=\kappa^2\bigg(\overset{em}{T}{}^2_{\;\;2}-H\overset{em}{T}{}^2_{\;\;0}\bigg)\;,
\end{eqnarray}
$\nu=3, i=3$
\begin{eqnarray}\label{ft33}
-\frac{D''}{D}f_T(T)-\frac{T'D'}{D}T'f_{TT}(T)+\frac{1}{2}f(T)=\kappa^2\overset{em}{T}{}^3_{\;\;3}\;.
\end{eqnarray}
Apparently, the non-symmetric components of the modified Einstein
equation are consistent with the tetrad anstaz given in
Eq.~(\ref{tet}). In the above equations, all other components of
$\overset{em}{T}{}^\mu_{\;\;\nu}$ must be zero, which means that,
$\overset{em}{T}{}_{\mu\nu}$, has the cylindrical symmetry  as
expected. In a G\"{o}del-type spacetime, the energy-momentum tensor
in a local basis, $\overset{em}{T}_{ab}$ given in~(\ref{Tab}), has a
general form: $\overset{em}{T}{}_{ab}=diag(\rho, p_1,p_2,p_3)$.
Using $\overset{em}{T}{}_{\mu\nu}=e^a_\mu e^b_\nu
\overset{em}{T}{}_{ab} $, we have
\begin{eqnarray}
\overset{em}{T}{}_{00}=\rho,\;\;\overset{em}{T}{}_{11}=p_1,\;\;\;\overset{em}{T}{}_{22}=H^2\rho+D^2p_2,\;\;\;\overset{em}{T}{}_{33}=p_3,\;\;\;\overset{em}{T}{}_{02}=\overset{em}{T}{}_{20}=H\rho\;.
\end{eqnarray}
One can then find easily
\begin{eqnarray}\label{costraint}
\overset{em}{T}{}^2_{\;\;0}=0,\;\;\;\;\overset{em}{T}{}^0_{\;\;2}=H\bigg(\overset{em}{T}{}^0_{\;\;0}-\overset{em}{T}{}^2_{\;\;2}\bigg)\;\;.
\end{eqnarray}
Thus, Eq.~(\ref{ft02}) seems to give an extra constraint on $f(T)$
gravity.  This equation is satisfied automatically in a
G\"{o}del-type spacetime, since $T$, as shown in Eq.~(\ref{TG}), is
a constant in a G\"{o}del-type universe.  Furthermore,  it is easy
to see  that, in a G\"{o}del-type spacetime, Eq.~(\ref{ft02}) gives
the same expression as Eq.~(\ref{ft22}). Four independent field
equations are obtained, which is consistent with the anstaz of
tetrad. In addition, one can check that the field equations (23-28)
for the vierbein given in (\ref{tet}) can also be obtained from an
action constructed by replacing the specific form of $T$ (\ref{T})
with the general action of $f(T)$ theory. Therefore,  the dynamical
equations are consistent, which means that the tetrad anstaz given
in Eq.~(\ref{tet}) is  a good guess for the G\"{o}del-type
spacetime.

\section{G\"{o}del-type universe in $f(T)$ theory}\label{ftgodel}

To show the possibility of  existence of the closed timelike curves and
the causality feature in $f(T)$ gravity, we consider the G\"{o}del-type metric,  which has the form of Eq.~(\ref{ds21})  with $H$ and $G$ being:
\begin{eqnarray}
H(r)=\frac{4\omega}{m^2}\sinh^2\bigg(\frac{mr}{2}\bigg)\;,
\end{eqnarray}
\begin{eqnarray}
G(r)=\frac{4}{m^2}\sinh^4\bigg(\frac{mr}{2}\bigg)\bigg[\coth^2\bigg(\frac{mr}{2}\bigg)-\frac{4\omega^2}{m^2}\bigg]\;,
\end{eqnarray}
where $\omega$ and $m$  ($-\infty< m^2<+\infty, 0<\omega^2$) are two constant parameters
used to classify different G\"{o}del-type geometries.
Thus, we have
\begin{eqnarray}
D(r)=\frac{1}{m}\sinh(mr)\;.
\end{eqnarray}

Substituting the expressions of $H$ and $D$ into Eq.~(\ref{T}), one can obtain easily
\begin{eqnarray}\label{TG}
T=2\omega^2\;,
\end{eqnarray}
which is a positive constant.

If $G(r)<0$, Eq.~(\ref{ds21}) shows that one type of  closed
timelike curve, called  noncausal G\"{o}del circle \cite{Godel},
exists in the case of $t, z, r=const$.  This means a violation of
causality. For a particular case of $0<m^2<4\omega^2$, the causality
violation region, i.e., $G(r)<0$ region, exists if
\begin{eqnarray}
\tanh^2\frac{mr}{2}<\frac{m^2}{4\omega^2}\;.
\end{eqnarray}\label{critcalr}
Thus, one can define a critical radius $r_c$~\cite{Godel, RebClif05e, Reb09b1, Reb10ed}
\begin{eqnarray}\label{crirad}
\tanh^2\frac{mr_c}{2}=\frac{m^2}{4\omega^2}\;,
\end{eqnarray}
beyond which, $G(r)<0$ and  causality is violated. When $m=0$, the critical radius is $r_c=1/\omega$.
When $m^2=4\omega^2$,  $r_c=+\infty$,
which means that a breakdown of causality is avoided.
Thus, the codomain range of $r_c$ is
$r_c \in(1/\omega, +\infty)$.   Therefore, the condition for non-violation of causality is $m^2\geq 4\omega^2$ or $r<r_c$.
For the case in which $m^2=-\mu^2<0$, both
$H(r)=\frac{4\omega}{\mu^2}\sin^2(\frac{\mu r}{2})$ and
$G(r)=\frac{4}{\mu^2}\sin^4(\frac{\mu r}{2})[\cot^2(\frac{\mu r}{2})-\frac{4\omega^2}{\mu^2}]$
are periodic functions. Thus, an infinite circulation of causal and noncausal ranges
appears~\cite{Reb09b1, Reb10ed}.

It is easy to see that, if one further defines a set of bases $\{\theta^a\}$:
\begin{eqnarray}
\theta^0=dt+H(r)d\phi,\qquad \theta^1=dr,
\end{eqnarray}
\begin{eqnarray}
\theta^2=D(r)d\phi,\qquad \theta^3=dz,
\end{eqnarray}
the Go\"{o}del-type line element can be simplified to be:
\begin{eqnarray}
ds^2=\eta_{ab}\theta^a\theta^b\;,
\end{eqnarray}
where $\eta_{ab}=diag(1,-1,-1,-1)$ is the Minkowski metric.
By choosing $\{\theta^a\}$ as  basis,
the $f(T)$ field equation~(\ref{motion2}) becomes:
\begin{eqnarray}\label{ds23}
A_{ab}f_T(T)+\frac{1}{2}\eta_{ab}f(T)=\kappa^2\overset{em}{T}_{ab}\;.
\end{eqnarray}
Here, both  $f(T)$ and $f_T(T)$ are evaluated at $T=2\omega^2$.   The second term of Eq.~(\ref{motion2}) is discarded in obtaining the   above   equation since the torsion scalar $T$ is a constant.
We find that the nonzero components of  $A_{ab}$ are
\begin{eqnarray}\label{A}
A_{00}=2\omega^2-m^2, \quad A_{11}=A_{22}=2\omega^2, \quad A_{33}=m^2\;.
\end{eqnarray}
Thus, we obtain a very simple form of  the field equation in $f(T)$ gravity, which will help us discuss the causality issue.

 \section{Causality Problem in $f(T)$ theory}\label{ftcondition}

One can see, from Eq.~(\ref{ds23}), that, in order to discuss the causality problem, the matter source is a very important component. As was obtained in \cite{RebClif05e, Reb09b1, Reb10ed}, different matter sources may lead  to different results. In this paper, we assume that the matter source consists of two different components: a perfect fluid and a scalar field.  Thus, the energy-momentum tensor $\overset{em}{
T}{}_{ab}$ has the form
\begin{eqnarray}\label{Tab}
\overset{em}{T}_{ab}=\overset{m}{T}_{ab}+\overset{s}{T}_{ab}\;,
\end{eqnarray}
where, $\overset{m}{T}_{ab}$ and $\overset{s}{T}_{ab}$ correspond to the energy-momentum tensors of  the
perfect-fluid and the scalar field,  respectively. In  basis $\{\theta^a\}$, $\overset{m}{T}_{ab}$ and $\overset{s}{T}_{ab}$ can be expressed as
\begin{eqnarray}
\overset{m}{T}_{ab}=(\rho+p)u_a u_b-p\eta_{ab}\;,
\end{eqnarray}
\begin{eqnarray}
\overset{s}{T}_{ab}=D_a\Phi D_b\Phi-\frac{1}{2}\eta_{ab}D_c\Phi
D_d\Phi\eta^{cd}\;,
\end{eqnarray}
where $u_a=(1,0,0,0)$, $\rho$ and $p$ are the energy density and
pressure of the perfect fluid, respectively, and $p=\text{w}\rho$
with $\text{w}$ being  the equation of state parameter. $\Phi$ is
the scalar field, and $D_a$ denotes the covariant derivative
relative to the local basis $\theta^a$.
The scalar field equation is
$\square \,\Phi = \eta^{ab}_{}\,\nabla_{a} \nabla_{b} \,\Phi\,=0$.  It
is easy to prove that
 $\Phi (z)= \varepsilon z + \text{const}$ with a constant amplitude $\varepsilon$  satisfies this field equation
~\cite{Reboucas1983}. Using the solution $\Phi (z)=\varepsilon z + \text{const}$,  one can obtain  the nonvanishing
components of $\overset{s}{T}_{ab}$
\begin{equation}  \label{S-comp}
\overset{s}T_{00} = - \overset{s}T_{11} = - \overset{s}T_{22} = \overset{s}T_{33} = \frac{\varepsilon^2}{2}\,,
\end{equation}
Thus, the energy-momentum
tensor of matter source becomes
\begin{eqnarray}
\overset{em}{T}_{ab}=diag\bigg(\rho+\frac{\varepsilon^2}{2}\;, \text{w}\rho-\frac{\varepsilon^2}{2
}\;, \text{w}\rho-\frac{\varepsilon^2}{2}\;, \text{w}\rho+\frac{\varepsilon^2}{2}\bigg)  \label{emt}\;.
\end{eqnarray}
Substituting Eqs.~(\ref{A}) and~(\ref{emt}) into the $f(T)$ field equation (Eq.~(\ref{ds23})), we find
\begin{eqnarray}\label{cfe1}
(2\omega^2-m^2)f_T(T)+\frac{1}{2}f(T)=\kappa^2(\rho+\frac{\varepsilon^2}{2})\;;
\end{eqnarray}
\begin{eqnarray}\label{cfe2}
2\omega^2f_T(T)-\frac{1}{2}f(T)=\kappa^2(\text{w}\rho-\frac{\varepsilon^2}{2})\;;
\end{eqnarray}
\begin{eqnarray}\label{cfe3}
m^2f_T(T)-\frac{1}{2}f(T)=\kappa^2(\text{w}\rho+\frac{\varepsilon^2}{2})\;.
\end{eqnarray}
Since the effective Newton gravity constant in $f(T)$ gravity becomes $G_{N,eff}=G_N/f_T(T)$~\cite{Zheng2011}, only the case $f_T(T)>0$ will be considered in the following in order to ensure a positive  $G_{N,eff}$.
From Eqs.~(\ref{cfe1}) and (\ref{cfe2}), one can derive  a  relation between $m$ and $\omega$:
\begin{eqnarray}
m^2=2\omega^2\bigg[1+\frac{\varepsilon^2}{\rho(1+\text{w})+\varepsilon^2}\bigg]\;,
\end{eqnarray}
which implies that  the critical radius of the G\"{o}del's circle, Eq.~(\ref{crirad}), satisfies
\begin{eqnarray}  \label{r_c}
\tanh^2\left(\frac{mr_c}{2}\right)=1-\frac{\rho(1+\text{w})}{2[\rho(1+\text{w})+\varepsilon^2]}\;.
\end{eqnarray}
Obviously, different  matter sources give rise to different critical
radii and therefore different causality structures, e.g. when
$\varepsilon\rightarrow0$, we have a finite $r_c$, while for
$\rho\rightarrow 0$, $r_c=\infty$. Therefore,  a violation of
causality may occur for the case of a perfect fluid as the matter
source, whereas causality is preserved in the case of a scalar
field. In order to show the causality feature in more detail and the
conditions for obtaining the G\"{o}del-type solutions, we will
divide our discussion into two special cases:
$\varepsilon\rightarrow0$ and $\rho\rightarrow0$. In addition, a
concrete $f(T)$ model will be considered.

\subsection{ $\varepsilon^2\rightarrow0$}\label{ftlimit1}

$\varepsilon^2\rightarrow0$ corresponds to the case that the universe only contains a perfect fluid.
Since $f_T(T)>0$, Eqs.~(\ref{cfe1}), (\ref{cfe2}), and (\ref{cfe3})
reduce to:
\begin{eqnarray}\label{mfe3}
m^2=2\omega^2\;;
\end{eqnarray}
\begin{eqnarray}\label{mfe2}
Tf_T(T)=\kappa^2\rho(1+\text{w})\;;
\end{eqnarray}
\begin{eqnarray}\label{mfe1}
f(T)=2\kappa^2\rho\;.
\end{eqnarray}
 From Eqs.~(\ref{mfe2}, \ref{mfe1}), it is easy to see that, in
the limit of general relativity without a cosmological constant
($f(T)=T$), $\text{w}=1$ is required to ensure the existence of the
 G\"{o}del-type solutions~\cite{godelnote1978, RebClif05e, Reb09b1, Reb10ed}. This means that a violation of causality in general relativity is
 only possible for the so-called stiff fluid ($\text{w}=1$) which is not a normal fluid in our Universe.
In $f(T)$ theory, $Tf_T(T)>0$ and $\rho>0$ lead to $\text{w}>-1$.
So,   the perfect fluid must satisfy the weak energy condition
($\rho>0$ and $\rho(1+\text{w})>0$). Using the above results, the
equation of state can be expressed as a function of  the torsion
scalar:
\begin{eqnarray}\label{wofft}
\text{w}=\frac{2Tf_T(T)}{f(T)}-1\;.
\end{eqnarray}
  Different from general relativity that requires $\text{w}= 1$
for  perfect-fluid G\"{o}del solutions, the equation of state
parameter of the fluid $\text{w}$ in $f(T)$ gravity can differ from
one and its value is determined by  concrete $f(T)$ models. For
example, a special $f(T)=\lambda T^{\delta}$ gives
$\text{w}=2\delta-1$,
from which one can see that $\text{w}$ can be an arbitrary number
for an arbitrary $\delta$. So, even normal matter, such as
pressureless matter or radiation, can lead to a violation of
causality in certain $f(T)$ theories. This indicates that  the
issue of causality violation seems more severe in $f(T)$ gravity
than in general relativity where only an exotic stiff fluid allows
the existence of G\"{o}del-type solutions.   From Eqs.~(\ref{mfe3}),
(\ref{mfe2}) and (\ref{mfe1}), and using $T=2\omega^2$, we find that
 the critical radius given in Eq.~(\ref{r_c}) becomes
\begin{eqnarray}
r_c=2\text{tanh}^{-1}\bigg(\frac{1}{\sqrt{2}}\bigg)\cdot\sqrt{\frac{f_T(T)}{(1+\text{w})\kappa^2\rho}}\;,
\end{eqnarray}
which is dependent both  on the specifics of $f(T)$ theory and the properties of the perfect fluid.

Now, let us consider  a concrete power law $f(T)$ model~\cite{Linder}
\begin{eqnarray}
f(T)=T-\alpha T_* \left(\frac{T}{T_*}\right) ^n\;,
\end{eqnarray}
where $\alpha$ and $n$ are model parameters, and $T_*$ is a special
value of the torsion scalar, which is introduced to make $\alpha$
dimensionless.   $|n|\ll 1$ is required in order to obtain an usual
early cosmic evolution~\cite{Wu2010b}.  The current cosmic
observations give that $\alpha=-0.79^{+0.35}_{-0.79}$ and
$n=0.04^{+0.22}_{-0.33}$ at the $68.3\%$ confidence
level~\cite{Wu2010a}. Thus, a negative $\alpha$ is favored by
observations.  In term of  Eq.~(\ref{wofft}), the equation of state
of the perfect fluid becomes
\begin{eqnarray}\label{wpow}\label{consofwpow1}
\text{w}=1-\frac{2\alpha(n-1)T^{1-n}_*}{T^{1-n}-\alpha T^{1-n}}\;.
\end{eqnarray}
The equation above can be re-expressed as
\begin{eqnarray}\label{consofwpow2}
\frac{\alpha(2n-1-\text{w})}{1-\text{w}}=\left(\frac{T}{T_*}\right)^{1-n}>0\;,
\end{eqnarray}
where a positive $T/T_*$ is considered.    Recalling  $\alpha<0$ and $\text{w}>-1$,  from Eq.~(\ref{consofwpow2}) one can obtain the possible ranges of $\text{w}$ for the G\"{o}del-type universes
\begin{eqnarray}\label{wpowrange}
1>\text{w}>-1+2n \quad(1>n>0)\;,\qquad 1>\text{w}>-1 \quad(n<0)\;.
\end{eqnarray}
For this power law model,  the critical radius has the form
\begin{eqnarray}\label{powrc}
r_c=2\left[\frac{\alpha(2n-1-\text{w})}{1-\text{w}}\right]^{\frac{1}{2(n-1)}}\text{tanh}^{-1}(1/\sqrt{2})\;,
\end{eqnarray}
which is determined completely by the model parameters and the equation of state of the perfect fluid.

\subsection{ $\protect\rho\rightarrow0$}\label{ftlimit2}

This is the case of  a scalar field as the matter source.
Eqs.~(\ref{cfe1}),~(\ref{cfe2}), and  (\ref{cfe3}) now reduce to
\begin{eqnarray}\label{sfe0}
m^2=4\omega^2\;,
\end{eqnarray}
\begin{eqnarray}\label{sfe1}
Tf_T(T)=\kappa^2\varepsilon^2\;,
\end{eqnarray}
\begin{eqnarray}\label{sfe2}
f(T)=3\kappa^2\varepsilon^2\;.
\end{eqnarray}
Note that (\ref{sfe1}) and (\ref{sfe2})
combined together admit a  relation between $T$ and $f(T)$:
  \begin{eqnarray}  \label{sig}
3Tf_T(T)-f(T)=0\;,
\end{eqnarray}
which constrains the class of solutions with no violation
of causality.
For the power law model,  the causal G\"{o}del-type solution gives that the torsion scalar should satisfy
\begin{eqnarray}\label{ssfpow}
T=2\omega^2=\left[-\frac{(1-3n)\alpha}{2}\right]^{\frac{1}{1-n}}T_*\;.
\end{eqnarray}
Thus, $n <1/3$ is required if the numerator  of $\frac{1}{1-n}$ is not even   since the observations show $\alpha<0$.

\section{Conclusions}\label{ftconclusion}
$f(T)$ theory, a new modified gravity, provides an alternative way
to explain the present accelerated cosmic acceleration with no need
of dark energy. Some problems, including large scale structure,
local Lorentz invariance, and so on, of this modified gravity have
been discussed. In this paper, we study the issue of causality in
$f(T)$ theory by examining the possibility of existence of the
closed timelike curves in the G\"{o}del metric. Assuming that the
matter source is a scalar field or a perfect fluid, we examine the
existence of the G\"{o}del-type solutions. For the scalar field
case, we find that $f(T)$ gravity allows a particular G\"{o}del-type
solution with $r_c\rightarrow\infty$, where $r_c$ is the critical
radius beyond which the causality is broken down.  Thus,    the
violation of causality can be forbidden.
In the case of a perfect fluid as the matter source,  we find
that the  fluid must have an equation of state parameter greater than minus one
and this parameter should satisfy Eq.~(\ref{wofft}) for the
G\"{o}del-type solutions to exist. For certain $f(T)$ models, the
perfect fluid that allows the G\"{o}del-type solutions  can even be
normal matter, such as pressureless matter or radiation. Since the
critical radius $r_c$ of perfect fluid G\"{o}del-type solutions
which  depends on both matter and gravity is finite, the issue of
causality violation seems more severe in $f(T)$ gravity than in
general relativity where only an exotic stiff fluid allows the
existence of G\"{o}del-type solutions.

\begin{acknowledgments}
 PXW would like to thank Prof.
Qingguo Huang for helpful discussions. This work was supported by
the National Natural Science Foundation of China under Grants Nos.
10935013, 11175093 and 11075083, Zhejiang Provincial Natural Science
Foundation of China under Grants Nos. Z6100077 and R6110518, the
FANEDD under Grant No. 200922, the National Basic Research Program
of China under Grant No. 2010CB832803, the NCET under Grant No.
09-0144, and K.C. Wong Magna Fund in Ningbo University.

\end{acknowledgments}

\end{document}